\documentclass[prd,twocolumn,showpacs,amsfonts]{revtex4}

\usepackage{epsfig}
\usepackage{graphics}
\usepackage{amsmath}

\newcommand{\be}{\begin{equation}}
\newcommand{\ee}{\end{equation}}

\begin{document}
\author{Jacob D. Bekenstein and Eva Sagi}
\affiliation{Racah Institute of Physics, Hebrew University of
Jerusalem, Jerusalem 91904, Israel\\}
\title{Do Newton's $G$ and Milgrom's $\mathfrak{a}_0$ vary with cosmological epoch ? }
\begin{abstract}
In the scalar tensor gravitational theories Newton's constant $G_N$ evolves in the expanding universe.  Likewise, it has been speculated that the acceleration scale $\mathfrak{a}_0$ in Milgrom's modified Newtonian dynamics (MOND) is tied to the scale of the cosmos, and must thus  evolve.  With the advent of relativistic implementations of the modified dynamics, one can address the issue of variability of the two gravitational ``constants'' with some confidence.  Using TeVeS, the Tensor-Vector-Scalar gravitational theory, as an implementation of MOND, we calculate the dependence of $G_N$ and  $\mathfrak{a}_0$ on the TeVeS parameters and the coeval cosmological value of its scalar field, $\phi_c$. We find that $G_N$, when expressed in atomic units, is strictly nonevolving,  a result fully consistent with recent empirical limits on the variation of $G_N$.   By contrast, we find that $\mathfrak{a}_0$ depends on $\phi_c$ and may thus vary with cosmological epoch. However, for the brand of TeVeS which seems most promising, $\mathfrak{a}_0$ variation occurs on a timescale much longer than Hubble's, and should be imperceptible back to redshift unity or even beyond it.  This is consistent with emergent data on the rotation curves of disk galaxies at significants redshifts.
\end{abstract}

\maketitle

\section{introduction}

The debate over the constancy of physical constants has simmered
ever since Dirac  ennunciated the Large Numbers
hypothesis: very large (or small) dimensionless universal
\textit{constants}  cannot occur in
the basic laws of physics~\cite{dirac}.   In particular, since the dimensionless gravitational constant is very small, the possibility of
variation of Newton's constant $G_N$ was raised.  The Brans-Dicke theory of
gravitation~\cite{brans-dicke}, among others, can describe such variation by adding to the Einstein-Hilbert action an action for a scalar field. 

The past few decades have witnessed extensive searches
for evidence of variation of some of the fundamental constants.  Among the
methods used have been astrophysical observations of the spectra of
distant quasars, searches for variations of planetary radii and
moments of inertia, investigations of orbital evolution, searches
for anomalous luminosities of faint stars, studies of abundance
ratios of radioactive nuclides, and (for current variations) 
laboratory intercomparison of precise clocks~\cite{Sisterna-Vucetich}.  To cite one example, current
data on elemental abundances, when compared with the theory of Big-Bang nucleosynthesis, limits the mean rate of  variation of
$G_N$ since early epochs to $(\dot{G_N}/G_N)<3\times 10^{-13} \mathrm{yr}^{-1}$
~\cite{copi:171301}. Obviously any proposed new theory of gravitation must be in harmony with this constraint.

$G_N$ is not the only gravity linked ``constant'' which might be variable.  Milgrom's  modified Newtonian dynamics-MOND
~\cite{Milgrom1983}, which was proposed to explain mass discrepancies in
galactic dynamics without calling on dark matter, introduces a new fundamental
parameter, $\mathfrak{a}_0$, with dimensions of acceleration.  In MOND, Newton's second law $\mathbf{a}=-\mathbf{\nabla}\Phi_N$ is replaced by
\be\label{mond_equation}
\tilde{\mu}\left(|\mathbf{a}|/\mathfrak{a}_0\right)\mathbf{a}=-\mathbf{\nabla}\Phi_N,
\ee 
where $\Phi_N$ is the usual Newtonian potential due to the baryonic matter alone, and the function $\tilde \mu(x)$ smoothly interpolates between
$\tilde\mu(x)=x$ at $x\ll 1$ and the Newtonian expectation  $\tilde
\mu(x)=1$ at $x\gg 1$. This phenomenological relation, with $\mathfrak{a}_{0}\simeq
10^{-10}\textrm{m}/\textrm{s}^2$, has had great success in
explaining the rotation curves of disk galaxies using only the
distribution of visible matter, as well as the slope and observed tightness of the Tully-Fisher relation, which correlates the luminosity (or baryonic mass) of a disk galaxy with its asymptotic rotational velocity~\cite{sand-ver}.  Recent reviews of MOND may be found in Refs. ~\onlinecite{McGaugh-Sanders,Bekenstein:2006,milgreview1,milgreview2}.

Use of MOND immediately raises a question: are the parameters $G_N$ and $\mathfrak{a}_0$ appearing in it constants of nature, or are they subject to spacetime changes, as is the $G_N$ in Brans-Dicke theory?

Milgrom~\cite{Milgrom1983} noticed that the observed value of  $\mathfrak{a}_0$ is quite close to $ cH_0$ where $H_0$ is the present epoch Hubble ``constant''.
He thus conjectured that $\mathfrak{a}_0$ may decrease together with the Hubble parameter on
cosmological timescales~\cite{Milgrom1983}.      By contrast Sanders~\cite{sanders2005} found that in the framework of the tensor-vector-scalar theory (BSTV) he proposed to provide a cosmological
basis for MOND, $\mathfrak{a}_0$  grows with time, and would differ significantly  at redshift $z\simeq 1$ from its present value.  This would imply  significantly reduced asymptotic rotational velocity for distant galaxies.  Sanders also remarked that in BSTV, just as in scalar-tensor theories, $G_N$  varies at a rate marginally in conflict with current observational bounds on $\dot G_N/G_N$. 

TeVeS, a new relativistic theory of gravity, was proposed by one of us~\cite{BekPRD} as a basis for MOND.  It  has been explored and subjected to a wide battery of tests~\cite{many}, and has also been extended in various directions~\cite{sanders2005,extend}.    TeVeS has MOND as its weak potential, low acceleration limit, while its weak potential, high acceleration limit is the usual Newtonian gravity.  TeVeS is endowed with three dynamical gravitational fields: a scalar field $\phi$, a
timelike unit normalized vector field $u_\alpha$, and the Einstein
metric $g_{\alpha\beta}$ on which the additional fields in the
theory propagate. The theory also employs a "physical" metric $\tilde{g}_{\alpha\beta}$ on which gauge, spinor and Higgs fields propagate. It is related to $g_{\alpha\beta}$ by
\be\label{grelation}
\tilde{g}_{\alpha\beta}=e^{-2\phi}g_{\alpha\beta}-2u_\alpha u_\beta
\sinh(2\phi). \ee 
The index of $u_\alpha$ or of $\phi_{,\alpha}$ is always raised with the metric $g^{\alpha\beta}$, the inverse of $g_{\alpha\beta}$.

The equations of motion for the fields in TeVeS derive
from a five-term action depending on four parameters: the fundamental gravity constant $G $, two dimensionless parameters $k$ and $ K$ and a fixed length scale $\ell$.  We use here the form of the action given in Ref.~\onlinecite{EvaBek}. Variation of the action with respect to $g^{\alpha\beta}$ yields the TeVeS
Einstein equations for $g_{\alpha\beta}$: \be\label{metric_eq}
G_{\alpha\beta}=8\pi G\left(
\tilde{T}_{\alpha\beta}+\left(1-e^{-4\phi}\right)u^\mu\tilde{T}_{\mu(\alpha}u_{\beta)}+\tau_{\alpha\beta}\right)+\theta_{\alpha\beta}.
\ee The sources here are the usual matter energy-momentum tensor
$\tilde T_{\alpha\beta}$ (related to the variational derivative of $S_m$ with respect to $\tilde g^{\alpha\beta}$), as well as  the energy-momentum tensors for the scalar and vector fields,
\begin{eqnarray}
\tau_{\alpha\beta}&\equiv& \frac{\mu(y)}{kG}\left(\phi_{,\,
\alpha}\phi_{, \,\beta}-u^\mu\phi_{, \mu}u_{(\alpha}\phi_{,
\,\beta)}\right)-\frac{\mathcal{F}(y)
 g_{\alpha\beta}}{2k^2 \ell^2 G}\,,
\\
\nonumber \theta_{\alpha\beta}&\equiv&
K\left(g^{\mu\nu}u_{[\mu,\,\alpha]}u_{[\nu,\,\beta]}-\frac{1}{4}g^{\sigma\tau}g^{\mu\nu}u_{[\sigma,\,\mu]}u_{[\tau,\,\nu]}g_{\alpha\beta}\right),
\\
&-&\lambda u_\alpha u_\beta
\end{eqnarray}
with \be \mu(y)\equiv\mathcal{F}'(y);\qquad  y\equiv kl^2 h^{\gamma\delta}
\phi_{,\,\gamma}\phi_{,\,\delta}. \ee Each choice of the function
$\mathcal{F}(y)$ defines a separate TeVeS theory. Its derivative $\mu(y)$ functions somewhat like the $\tilde{\mu}$ function in MOND. For $y>0$, $\mu(y)\simeq 1$ corresponds to the high acceleration, i.e., Newtonian, limit, while the limit $0<\mu(y)\ll 1$ corresponds to the deep MOND regime.  We shall only consider functions such that $\mathcal{F}>0$ and $\mu>0$ for either positive or negative arguments.

The equations of motion for the vector and scalar fields are obtained by varying the action with respect to $\phi$ and $u_\alpha$, respectively.  We have
\be
\left[\mu(y) h^{\alpha\beta}\phi_{,\,\alpha}\right]_{;\,\beta}
=kG\left[g^{\alpha\beta}+\left(1+e^{-4\phi}\right)u^\alpha
u^\beta\right] \tilde{T}_{\alpha\beta}\,, \label{scalar_eq}
\ee
for the scalar and 
\begin{eqnarray}
&&u^{[\alpha;\beta]}\;_{;\beta}+\lambda u^\alpha+\frac{8\pi}{k}\mu
u^\beta\phi_{,\,\beta}g^{\alpha\gamma}\phi_{,\,\gamma}\nonumber
\\
&&=8\pi
G\left(1-e^{-4\phi}\right)g^{\alpha\nu}u^\beta\tilde{T}_{\nu\beta}\,.
\label{vector_eq}
\end{eqnarray}
for the vector.   Additionally, there is the normalization condition on the vector
field: \be \label{normalization} u^\alpha
u_\alpha=g_{\alpha\beta}\,u^\alpha u^\beta=-1. \ee
The $\lambda$ in Eq.~(\ref{vector_eq}), the lagrange multiplier charged with the enforcement of the normalization condition, can be calculated from the vector equation. 

The three parameters, $k, K$ and $\ell$, all specific to TeVeS, are constant in the framework of the theory, as is $G$, the fundamental gravitational coupling constant, (which need not coincide with Newton's $G_N$).   As shown in Ref.~\onlinecite{BekPRD}, the measurable quantities $G_N$ and $\mathfrak{a}_0$ can be expressed in terms of $k, K, \ell$ and $G$.  However, that calculation of $G_N$ neglected the nonzero cosmological value $\phi_c$ of the scalar field in the scalar equation's matter source, thus obtaining spurious cosmological evolution of $G_N$. In this paper we carry out the calculations in great detail, and show
that TeVeS predicts a strictly nonvarying $G_N$ and only a weak cosmological evolution of $\mathfrak{a}_0$.  This is in full agreement with available observational constraints on the evolution of $G_N$, as well as the emerging constraints on evolution of $\mathfrak{a}_0$.  As extended rotation curves of high-$z$ galaxies become available in the future, they will make possible  a serious check of the TeVeS's prediction that $\mathfrak{a}_0$ evolves weakly.

In Sec.~\ref{sec:dim} we clarify the sense in which $G_N$ turns out to be constant by contrasting physical (atomic) units of length with the Einstein units in which the gravitational action looks simple.  In Sec~\ref{sec:Newton} we work in the Newtonian (strong acceleration-weak potential) limit of TeVes to calculate $G_N$ in terms of the fundamental constant $G$ and the TeVeS parameters $K$ and $k$.  Passing to the weak acceleration limit of TeVes we calculate in Sec.~\ref{sec:mond}  $\mathfrak{a}_0$ in terms of  the TeVeS parameters $\ell$, $K$ and $k$ and the cosmological value $\phi_c$.   In Sec.~\ref{sec:evolution} we estimate the cosmological evolution of $\mathfrak{a}_0$, first naively by assuming that it is tied to that  of the Hubble parameter which is taken to evolve \textit{a la} GR, and then by setting a bound on the rate of $\phi$ evolution from TeVeS's equations.  The latter method clearly shows that $\mathfrak{a}_0$ evolves slowly on Hubble's scale.  Sec.~\ref{conc} summarizes our conclusions.  In the Appendix we check our methodology by recovering the accepted $G_N$ for the Brans-Dicke gravitational theory.  

Greek indeces run over $0,1,2,3$ with $x^0=t$ representing time; a partial derivative with respect to $t$  is denoted by an overdot.  We set $c$ to unity everywhere.

\section{Dimensions and Units}
\label{sec:dim}

What does it mean to say that $G_N$ is not evolving?   After all, a \textit{dimensionfull} quantity can be caused to be constant in spacetime  by simply choosing the unit in which it is measured to have suitable spacetime variation~\cite{dicke}. So the only operationally meaningful statement of constancy of $G_N$ is that some \textit{dimensionless} combination of physical parameters involving $G_N$ does not evolve.  Let us thus specify such a constant combination in TeVeS. 

Following Dicke's masterful critique~\cite{dicke}, we choose in the present section to regard the metric coefficients as carrying dimensions of squared length and to think of all the coordinates themselves as dimensionless.  As mentioned, in TeVeS the equations for the material fields (spinor, gauge and Higgs fields) take their usual form when written on the physical metric $\tilde g_{\alpha\beta}$. In particular, we assume that in the stated formulation the \textit{dimensionless} gauge coupling constants and all elementary particle masses are constant in spacetime.  What is being assumed is that  the system of units reflected by $\tilde g_{\alpha\beta}$ uses a particle mass, say the proton's $m_p$, as its local mass unit.  Likewise, the physical parameters $c$ and $\hbar$, which appear in the various equations, are supposed constant.  Since $\hbar, c$ and $m_p$ all bear different dimensions,  this last requirement has fixed the system of units (up to the trivial freedom to double the length unit everywhere, etc.).  The statement that $G_N$ is nonevolving in such ``atomic'' units is thus equivalent to the statement that $G_N m_p^2/\hbar c$ is a spacetime constant.

The above units differ from those carried by the Einstein metric $g_{\alpha\beta}$ which is the one used in formulating the TeVeS equations.  (Here it may prove conceptually useful to regard the $\tilde T_{\alpha\beta}$, which is calculated by varying the matter action with respect to $\tilde g^{\alpha\beta}$, as reexpressed in terms of $g_{\alpha\beta}$).   According to Eq.~(\ref{grelation}), for like coordinate increments, the \textit{physical} distance in the space orthogonal to $u^\alpha$ (the space whose metric is $g_{\alpha\beta}+u_\alpha u_\beta$)  is a factor $e^{-\phi}$ times the distance paced out by the Einstein metric itself.  By contrast, the physical distance collinear with the timelike vector field $u^\alpha$ is  $e^\phi$ times that given by the Einstein metric.  In other words, the Einstein unit of length is not only spacetime varying but also spacetime anisotropic \textit{with respect to that in ``atomic'' units}.  In Einstein units we may still regard  $c$ and $\hbar$ (but not $m_p$ or its corresponding Compton length) as constants.  Accordingly, we have set $c=1$ everywhere.  As mentioned $G$ is constant in Einstein units.  Were we to set it as well as $\hbar$ to unity, we would be in Planck units.  However, we shall refrain from this last step and continue to exhibit $G$ explicitly.  The main question we shall be asking is, how do $G_N$ or $\mathfrak{a}_0$, when appropriately calculated in physical units, relate to the TeVeS constants $G$, $\ell$, etc.?

\section{Newton's $G_N$ is constant}
\label{sec:Newton}

We begin by showing the relation between Newton's constant and the
TeVeS coupling constant $G$. By definition, Newton's constant
enters through the relation 
\be\label{NewtPot}
\Phi_N=-G_N\int\frac{\tilde{\rho}(\tilde{\mathbf{x}}')}{|\tilde{\mathbf{x}}'-\tilde{\mathbf{x}}|}d^3\tilde{x}'
\ee with $m=\int\tilde{\rho}(\tilde{\mathbf{x}}')d^3\tilde{x}'$ the
physical mass and $\tilde{\mathbf{x}}$ the Cartesian coordinate that marks physical distance. Since
we expect TeVeS to have the customary weak field limit, the
Newtonian potential (\ref{NewtPot}) should enter in the customary way
in the linearized form of the physical metric (the metric measured by instruments made of matter). We thus
expect that
\be
\label{linear}
d\tilde{s}^2=-(1+2\Phi_N)d\tau^2+(1-2\Phi_N)d\tilde{\mathbf{x}}\cdot
d\tilde{\mathbf{x}}.
\ee
where $\tau$ is the  coordinate that marks physical time.

To be able to compare  Einstein and physical metrics, we must use the same coordinates for both.  To maintain consistency with
previous work ~\cite{BekPRD,EvaBek} we choose new coordinates $t$ and $\mathbf{x}$  in terms of which the \textit{asymptotic} physical metric, though flat, differs slightly from standard Minkowsky form. The relations between physical distance and time, $\tilde{\mathbf{x}}$ and $\tau$, and the coordinates ${\mathbf{x}}$ and $t$ are
\be\tilde{\mathbf{x}}=e^{-\phi_c}{\mathbf{x}}\label{coordrel};\qquad \tau=e^{\phi_c} t.
\ee 
With this in mind, we
can rewrite the weak field limit of the physical metric outside  its
source in the form
\be\label{LinearizedMetric}
d\tilde{s}^2=-(1+2\Phi_N)e^{2\phi_c}dt^2+(1-2\Phi_N)e^{-2\phi_c}d{\mathbf{x}}\cdot
d{\mathbf{x}}.
\ee  

To relate $G_N$ to $G$ we must solve TeVeS equations and use the solutions to construct the physical metric, which should turn out to be identical to
Eq.~(\ref{LinearizedMetric}).
Consider the following ansatz for the Einstein metric to first order:
\be\label{linearEE}
ds^2=-(1+2V)dt^2+(1-2V)d{\mathbf{x}}\cdot
d{\mathbf{x}}. \ee 
The vector field $u^\alpha$ must be timelike; we shall take our coordinate system to coincide with the ``rest frame'' established by $u^\alpha$.   In view of the normalization condition (\ref{normalization}), the vector field is given to
first order by 
\be
\label{linearizedu}u_\alpha=\{-(1+V),0,0,0\}. \ee The scalar field
may be written
 \be\phi=\phi_c+\delta\phi\label{linearizedphi},
\ee with $\phi_c$
the nonzero cosmological value of $\phi$ and $\delta\phi$ representing the local departure from it; $\delta\phi$ is to be regarded as of the
same order of smallness as $V$.

Now for the details.   To find the potential $V$, we need to solve the $G_{tt}$
Einstein equation to first order. We take for the energy-momentum
tensor the familiar ideal fluid form \be
\label{matterT}
\tilde{T}_{\alpha\beta}=\tilde{\rho}v_\alpha v_\beta
+\tilde{p}\,(\tilde{g}_{\alpha\beta}+v_\alpha v_\beta), \ee where
$\tilde{\rho}$ is the proper energy density, $\tilde{p}$ the
pressure and $v_\alpha$ the 4-velocity, all three referred to the physical metric. We assume that the fluid is stationary in the coordinates chosen, so the spatial part of $v^\alpha$ must vanish, i.e.,  $v_{\alpha}$ must be parallel to
$u_\alpha$.  When account is taken of $v_\alpha$ normalization with reference to
$\tilde{g}_{\alpha\beta}$ we get~\cite{BekPRD}
\be
v_{\alpha}=e^{\phi}u_{\alpha}.  \ee 
In the nonrelativistic approximation $\tilde{p}$ is negligible
relative to $\tilde{\rho}$; thus \be
\tilde{T}_{\alpha\beta}=\tilde{\rho}e^{2\phi}u_{\alpha}u_{\beta}.\label{linearizedEMtensor}
\ee

We now substitute (\ref{linearEE}),(\ref{linearizedu}),
(\ref{linearizedphi}), and (\ref{linearizedEMtensor}) into the TeVeS
equations, and retain only first order in $V$ and $\delta\phi$. We start by solving
the temporal component of the vector equation (\ref{vector_eq}) for $\lambda$ (the
other components are zero to linear order), obtaining
\be \lambda=-K\nabla^2 V-16\pi G\tilde{\rho}\sinh{(2\phi_c)}. \ee 
When this is substituted in the $G_{tt}$ Einstein equation, it becomes, to first order,
 \be \nabla^2 V=\frac{4\pi
G}{1-K/2}\tilde{\rho}e^{-2\phi_c}. \ee 
We have neglected in the r.h.s. terms of order $\delta\phi$ which would source terms of second order in $V$.  With the boundary condition $V\rightarrow 0$ for $|\mathbf{x}|\rightarrow\infty$ the solution is
\be\label{TeVeSV}
V=-\frac{e^{-2\phi_c}G}{1-K/2}\int\frac{\tilde{\rho}(\mathbf{x}')}{|\mathbf{x}'-\mathbf{x}|}d^3x'.
\ee 
in which integral $\tilde\rho$ is regarded as a function of the coordinates $\mathbf{x}$, not of the physical distances.

This result must make its way into the physical metric. Using transformation~(\ref{grelation})  in Eq.~(\ref{linearEE}) we get
 \be
d\tilde{s}^2=-(1+2V)e^{2\phi}dt^2+(1-2V)e^{-2\phi}d\mathbf{x}\cdot
d\mathbf{x}, \ee 
which becomes--to lowest order in $\delta\phi$-- 
\be\label{LinearizedTeVeSMetric}
d\tilde{s}^2=-(1+2V+2\delta\phi)e^{2\phi_c}dt^2+(1-2V-2\delta\phi)e^{-2\phi_c}d\mathbf{x}\cdot
d\mathbf{x}. \ee 
Comparing with Eq.~(\ref{LinearizedMetric}) we may identify the Newtonian potential
\be
\label{PhiN1}
\Phi_N=V+\delta\phi.
\ee
We now need only calculate $\delta\phi$.

In the scalar equation (\ref{scalar_eq}) we substitute $\mu=1$ because we are concerned with the Newtonian limit in which $\mu$ approaches unity; any small corrections to it may be discarded since we work here to first order in
$\delta\phi$: 
\be\label{scalareq_firstorder}
\nabla^2\delta\phi=kGe^{-2\phi_c}\tilde{\rho}. \ee
 Note the factor $e^{-2\phi_c}$, the correct asymptotic value of
$e^{-2\phi}$, which was missed in Eq.(53) of
Ref.~\onlinecite{BekPRD}.  The solution of this last equation in accordance with the boundary condition $\phi\rightarrow\phi_c$ as $\mathbf{x}\rightarrow\infty$ is  \be\label{deltaphi}
\delta\phi=-\frac{kGe^{-2\phi_c}}{4\pi}\int\frac{\tilde{\rho}(\mathbf{x}')}{|\mathbf{x}'-\mathbf{x}|}d^3x'.
\ee 

Substituting Eqs.~(\ref{TeVeSV}) and (\ref{deltaphi}) into Eq.~(\ref{PhiN1}) we get \be
\Phi_N({\bf x})=-\left(\frac{(2-K)k+8\pi}{(2-K)k}\right)e^{-2\phi_c}G\int\frac{\tilde{\rho}(\mathbf{x}')}{|\mathbf{x}'-\mathbf{x}|}d^3x'.
\ee 
 Finally, we use relation (\ref{coordrel}) to switch back to
physical length coordinates $\tilde{\bf x}$:
 \be\label{PhiN}
\Phi_N(\tilde{\bf x})=-\left(\frac{(2-K)k+8\pi}{(2-K)k}\right)G\int\frac{\tilde{\rho}(\tilde{\mathbf{x}}')}{|\tilde{\mathbf{x}}'-\tilde{\mathbf{x}}|}d^3\tilde{x}',\ee
where $\tilde\rho(\tilde{\bf x})\equiv \tilde\rho({\bf x}e^{-\phi_c})$ is the energy density distribution in physical units.
Comparing with Eq.~(\ref{NewtPot}) we obtain $G_N$ in terms of the TeVeS parameters:
\be G_N=\left(\frac{(2-K)k+8\pi}{(2-K)k}\right)G.
\label{GNfinal}
 \ee 

Thus the ratio $G/G_N$ turns out not to depend on $\phi_c$, the asymptotic cosmological value of $\phi$, a cosmologically evolving quantity.  Since $G, k$ and $K$ are constant parameters, $G_N$  (as measured in physical units)  does not evolve with cosmological epoch.   Further, since $G_N$ is observationally positive, while it is natural to expect that $G>0$, we must restrict $K$ to the ranges $K<2$ or $K>2+8\pi/k$.   It is amusing that one can reconcile the small observed value of $G_N$  (more properly of $G_N m_p^2/\hbar$) with strong gravity ($G m_p^2/\hbar$ which is not especially small) for the family of TeVeS theories for which $K$ is only very slightly above the critical value $2+8\pi/k$.  In such a theory the true Planck length $(G\hbar)^{1/2}$ could be commensurate with elementary particle scales like $\hbar/m_p$ or be even larger, all this without resorting to brane physics.

The result that $G_N$ is constant is surprising in view of the fact that TeVeS contains a scalar sector.  To check our methodology we work out,  in Appendix A, $G_N$ for Brans-Dicke theory  by following the track set out in the present section.  We obtain the accepted law of evolution.

\section{the mond acceleration scale}
\label{sec:mond} 

The MOND acceleration scale $\mathfrak{a}_0$ can also be calculated in terms
of the TeVeS parameters. Since MOND is the small
acceleration, weak potential limit of TeVeS~\cite{BekPRD}, we can again work with the linear approximation to the physical and Einstein metrics; however,
in the present case $\mu \ll 1$. As in Ref.~\onlinecite{BekPRD} we shall  look for a MOND-like equation of the form (\ref{mond_equation}).  We shall then attempt to identify the MOND function $\tilde{\mu}$ and the combination of TeVeS coupling constants which is equivalent to $\mathfrak{a}_0$. For simplicity, we shall assume spherical symmetry; it can be shown that our results hold for asymmetric systems as well~\cite{BekPRD}.

We write the Einstein metric for weak potentials exactly as in Eq.~(\ref{linearEE}), while in contrast to Eq.~(\ref{LinearizedMetric}) the physical metric is expected to be
\be\label{LinearizedMondMetric}
d\tilde{s}^2=-(1+2\Phi)e^{2\phi_c}dt^2+(1-2\Phi)e^{-2\phi_c}d\mathbf{x}\cdot
d\mathbf{x}, \ee 
with $\Phi$, the MOND gravitational potential, replacing $\Phi_N$. By transforming from Einstein to physical metric in accordance with Eqs.~(\ref{grelation}) and (\ref{linearizedu})-(\ref{linearizedphi}) we find to first order that  
\be\label{newPhi}
\Phi=V+\delta\phi.
\ee  The contribution of $V$ to $\Phi$ is the same as that to $\Phi_N$ because the terms depending on $\mu$ in the $G_{tt}$ equation from which $V$ arises are all of second order in $\delta\phi$, and thus stand for higher order corrections. Thus we have
$V$ as in Eq.~(\ref{TeVeSV}).

In determining $\delta\phi$ here we must
take into account the fact that in the weak acceleration limit $\mu<1$.  The scalar equation (\ref{scalar_eq}) takes the form \be\label{deltaphiMond} \mathbf{\nabla}\cdot
\left[\mu\left(k\ell^2\left(\mathbf{\nabla}\delta\phi\right)^2\right)\mathbf{\nabla}\delta\phi\right]=kGe^{-2\phi_c}\tilde{\rho}.\ee
Comparing this equation with Poisson's and using Gauss' theorem in the spherically symmetric case gives
\be\label{nabla_delta_phi_Mond}
\nabla\delta\phi=-\frac{kGe^{-2\phi_c}}{4\pi \mu}\nabla\int\frac{\tilde{\rho}(\mathbf{x}')}{|\mathbf{x}'-\mathbf{x}|}d^3x'.
\ee 
  Then in view of Eqs.~(\ref{TeVeSV}) and (\ref{newPhi})  the gradient of the total potential $\Phi$ satisfies \begin{eqnarray}\label{Phieq} \tilde{\mu}\nabla\Phi&=&-Ge^{-2\phi_c}\nabla\int\frac{\tilde{\rho}(\mathbf{x}')}{|\mathbf{x}'-\mathbf{x}|}d^3x';\\
\label{mutilde} \tilde{\mu}&\equiv& 
\left[\left(\frac{1}{1-K/2}+\frac{k}{4\pi\mu}\right)\right]^{-1}.\end{eqnarray}

Eq.(\ref{Phieq}) is the desired MOND-like equation.   To find Milgrom's parameter $\mathfrak{a}_0$ we proceed to the extreme MOND regime defined by the condition $\mu\ll k/(4\pi)$. There Eq.~(\ref{mutilde}) gives $\tilde\mu\approx 4\pi\mu/k $.  Substituting this in Eq.~(\ref{Phieq}) and comparing the result with Eq.~(\ref{nabla_delta_phi_Mond})  reveals that in the said limit $\nabla\Phi\approx\nabla\delta\phi$.  This implies that in  Eq.~(\ref{PhiN1}) $\nabla V$ is then negligible. 

Instead of focusing on the toy form of $\mu(y)$ from Ref.~\onlinecite{BekPRD}, or any other ansatz for it, let us be very general.  We just require that for $0<y\ll 1$, $\mu(y)\approx D\surd y$ where $D$ is some positive constant.   Going to sufficiently small $y$ so that $\mu\ll k/(4\pi)$ we have, in view of the last paragraph, that
\be
\tilde\mu\approx 4\pi\mu/k\approx 4\pi D k^{-1/2}\ell|\nabla\delta\phi| \approx 4\pi D k^{-1/2}\ell|\nabla\Phi|.
\ee 
Substituting this in Eq.~(\ref{Phieq}) and transforming all ocurrences of ${\bf x}$ (including in the gradients)  to $\tilde{\bf x}$  by means of Eq.~(\ref{coordrel}), we get  the extreme MOND equation~\cite{Milgrom1983},
\be
|\nabla_{\tilde{\bf x}}\Phi|\nabla_{\tilde{\bf x}}\Phi/\mathfrak{a}_0=\nabla_{\tilde{\bf x}}\Phi_N, 
\ee
with
\be
\label{a0}
\mathfrak{a}_0={G\over G_N}{\surd k\, e^{\phi_c}\over 4\pi D\ell}.
\ee
We have here employed the definition (\ref{NewtPot}); it is understood that $\Phi$ also is to be regarded as a function of the physical length coordinates $\tilde{\bf x}$.  We see that Milgrom's acceleration scale $\mathfrak{a}_0$ depends on the TeVeS parameters $k, K$ and $\ell$ (all constant in Einstein units) as well as on the constant coefficient $D$ associated with the function ${\cal F}(y)$.  But unlike $G_N$,  $\mathfrak{a}_0$ is predicted in TeVeS to evolve cosmologically in consonance with $e^{\phi_c}$.   How fast an evolution it is capable of is the subject of the next section.

\section{Predicted evolution of $\mathfrak{a}_0$}
\label{sec:evolution}
\subsection{The naive MOND viewpoint}

The pure MOND paradigm is ambiguous about the time evolution of $\mathfrak{a}_0$.  Milgrom~\cite{Milgrom1983,milgreview1,milgreview2} remarks on the numerical coincidence between $\mathfrak{a}_0$  and the observed $c H_0$   ($H_0$ is the present value of the Hubble parameter $H$) or $\mathfrak{a}_0$ and the value of the cosmological constant $\Lambda$ inferred from the acceleration of the cosmos.   If the former coincidence bespeaks of a physical connection, then one would expect cosmological evolution of  $\mathfrak{a}_0$ with $\mathfrak{a}_0\propto H$,  while if it is the second coincidence that properly reflects the physics, then  $\mathfrak{a}_0$ should be strictly constant.   

How big an evolution to expect in the first case ?  Since naive MOND does not provide a consistent cosmology, we shall here use cosmology \textit{a la} GR.  Let us write the Friedmann equation in GR for a cosmological model with curvature index $\kappa$:
\be
\label{Friedmann}
H^2={\dot b^2\over b^2}=-{\kappa\over b^2}+{\Lambda\over 3}+{8\pi G\rho_{m0}  b_0^3 \over 3 b^3}.
\ee
Here $b=b(t)$ is the expansion factor with value $b_0$ at the present time, $\rho_{m0}$ is the present value of the mass density of pressureless matter, and  we are neglecting radiation's contribution because we focus on the more recent universe. Differentiating with respect to $t$ and dividing out by $2H$ gives
\be
\label{rate}
{\dot H\over H}=-\left({-\kappa\over H^2 b^2}+{4\pi G\rho_{m0}b_0^3 \over H^2 b^3} \right)H.
\ee

As customary we may introduce densities as fractions of the present critical density
\be
\Omega_{m}\equiv {8\pi G\rho_{m0} \over 3 H_0^2};\quad \Omega_\kappa\equiv {-\kappa\over H_0^2 b_0^2 }; \quad \Omega_\Lambda\equiv {\Lambda\over 3H_0^2},
\ee
so that $\Omega_m+\Omega_\kappa+\Omega_\Lambda=1$ on account of Eq.~(\ref{Friedmann}) evaluated at the present epoch (when $b=b_0$ and $H=H_0$).  We see that
\be
\label{rate}
(\dot H/H)_0=-(\Omega_\kappa+{\scriptstyle 3\over\scriptstyle 2}\Omega_m)H_0.
\ee
The standard cosmological model obtains values for the $\Omega$'s from various observations, e.g., those of the cosmological microwave background anisotropy spectrum.  $\Omega_\kappa$ comes out either zero (flat space) or positive (hyperbolic space) and very small on scale unity.  By contrast $\Omega_m$, which includes the contribution from putative dark matter, is assigned a value of about 0.25.   We may thus conclude that  \textit{at present} $\dot {\mathfrak{a}}_0/\mathfrak{a}_0$, which is the same as $(\dot H/H)_0$, should be about  $-0.25 H_0$.   Thus the present day timescale of $\mathfrak{a}_0$ variation is four times longer than the Hubble scale.   

As we go back in time $\mathfrak{a}_0$ should scale proportionately to the coeval $H$.  We may recast Eq.~(\ref{Friedmann}) as
\be
\label{Hz}
H=H_0\Big[\Omega_\kappa (1+z)^2+\Omega_\Lambda+\Omega_m(1+z)^3\Big]^{1/2}
\ee
with $1+z=b_0/b$.   With $\Omega_\kappa\approx 0$, $\Omega_m\approx 0.25$ and $\Omega_\Lambda\approx 0.75$ as in the standard model, the curvature term in the square brackets in the last equation remains negligible, while by $z\approx 1$ the matter term will have come to dominate the $\Lambda$ term.  We  then have for $z>1$
\be
\label{aMilg}
\mathfrak{a}_0(z)\approx \mathfrak{a}(0)(1+z)^{3/2}
\ee
which implies a drastic change of $\mathfrak{a}_0$ between $z$ of a few and today.

However, it could be claimed that  to keep in the spirit  of the MOND paradigm one should, apart from retaining $\Omega_\kappa\approx 0$, equate $\Omega_m$ with the baryon fraction $\Omega_b=0.04$ inferred in standard cosmology.   This last can easily accommodate still unobserved massive neutrino or baryonic matter which is nowadays invoked in MOND in connection with the large clusters of galaxies~\cite{sand-neutrino,point,milgreview1}.  Of course to be consistent we should then put $\Omega_\Lambda\approx 0.95$. With this set up the matter term in Eq.~(\ref{Hz})  becomes comparable with the $\Lambda$ term only for $z\approx 2$, and $\mathfrak{a}_0$ will follow the law (\ref{aMilg}) for  $z>2$.  For $z\ll 1$ we would have from Eq.~(\ref{rate}) that $\mathfrak{a}_0$ changes on a timescale 16 times longer than $H_0^{-1}$.

The above discussion is instructive; but it is hardly trustworthy as underlined by the contrasting results it can yield.  The crux of the problem is, of course, that MOND is not a nonrelativistic limit of GR, yet this last is being used to work out the cosmology.  This inconsistency can be avoided by calculating the cosmological evolution of $\mathfrak{a}_0$ entirely within TeVeS, which does have MOND as a nonrelativistic limit.

\subsection{The TeVeS viewpoint}

We found in Eq.~(\ref{a0}) that $\mathfrak{a}_0$, as defined by small scale MOND dynamics, has a $e^\phi$ dependence, where by $\phi$ is meant $\phi$'s cosmological value $\phi_c$.  We are thus invited to establish the cosmological evolution of $\phi$.  It will be useful to distinguish here, as we did in Sec.~\ref{sec:Newton}, between the coordinate time $t$ and the physical time $\tau$. 
 
The Einstein metric for a Friedmann-Robertson-Walker model is
\be
ds^2=-dt^2+b(t)^2[d\chi^2+f(\chi)^2 (d\theta^2+\sin^2\theta\, d\varphi^2) ]
\ee
where $f(\chi)$ is either $\chi$ (open model with flat spaces, $\kappa=0$) or $\sinh\chi$ (open model with hyperbolic spaces, $\kappa=-1$).  As in Sec.~VII of Ref.~\onlinecite{BekPRD} we shall take $u^\alpha=\{1,0,0,0\}$ and $\phi=\phi(t)$, consistent with the timelike character of the vector and the assumed isotropy and homogeneity of space.  Then according to Eq.~(\ref{grelation}) we obtain the physical line element $d\tilde s^2$ by multiplying the temporal part of $g_{\alpha\beta}$ by $e^{2\phi}$ and the spatial parts by $e^{-2\phi}$:
\be
d\tilde s^2=-d\tau^2+\tilde b(t)^2[d\chi^2+f(\chi)^2 (d\theta^2+\sin^2\theta\, d\varphi^2) ],
\ee
with
\be
\label{ta}
\tilde b=e^{-\phi} b; \qquad d\tau=e^{\phi} dt.
\ee
The $\tau$ here is the physical time since it acts as proper time of commoving observers, c.f. Eq.~(\ref{coordrel}).

From $\mathfrak{a}_0\propto e^\phi$ it follows that
\be
\label{a0dot}
{d\mathfrak{a}_0/d\tau\over \mathfrak{a}_0} = {d\phi\over d\tau}
\ee
The first integral of Eq.~(\ref{scalar_eq})  for $\phi$ is given in Ref.~\onlinecite{BekPRD} for the case of ideal fluid matter:
 \be
 \label{first}
 \mu(-2k\ell^2\dot\phi^2)\dot\phi={-k\over 2 b^3}\int_0^t G(\tilde
\rho+3\tilde p)e^{-2\phi}b^3 dt,
\ee
Here, as earlier, an overdot designates a derivative with respect to $t$, not $\tau$. Since the physical energy density, $\tilde\rho$, the physical pressure, $\tilde p$, and the TeVeS parameter $k$ are all positive, we see that $\dot\phi$, and consequently also $d\phi/d\tau$, are negative.  Thus by Eqs.~(\ref{ta})-(\ref{a0dot}) $\mathfrak{a}_0$ is strictly decreasing with physical time $\tau$.  But because the integral above includes contributions from early times when $\tilde p$ is not negligible, and because of the complicated factor $\mu$, the said equation is far from convenient for estimating $\dot\phi$.  We shall instead estimate $d\phi/d\tau$ by way of the Einstein equations.

First we compute the physical Hubble parameter $\tilde H$:
\be
\label{tildeH}
\tilde H\equiv {d\tilde b/d\tau\over \tilde b}=e^{-\phi}{\dot b\over b}-{d\phi\over d\tau}.
\ee
  Next we compute $\lambda$  from the vector equation (\ref{vector_eq}); it takes the form~\cite{BekPRD}
\be
\lambda=8\pi\big[\mu\dot\phi^2/k-2G\sinh(2\phi)\tilde\rho\big].
\ee
Finally  we write down Einstein's equations (\ref{metric_eq}) \textit{sans} the cosmological constant 
and with a perfect fluid as matter,
  \be
 {\dot b^2\over b^{2}}=-{\kappa\over b^2}+{8\pi G\over 3}\tilde\rho
e^{-2\phi}+{16\pi \mu(y) \dot\phi^2\over 3k}
+ {4\pi\mathcal{F}(y)\over 3k^2\ell^2},
\label{Friedmann2}
\ee
where $y$ here is identical to the argument of $\mu$ in Eq.~(\ref{first}).   Taking into account that $\mathcal{F}>0$ and $\mu>0$ and that $\dot\phi<0$, we see that for a spatially flat or hyperbolic cosmological model  ($\kappa\leq 0$)
\be
{\dot b\over b}>-\left({16\pi\mu\over 3k}\right)^{1/2} e^\phi {d\phi\over d\tau}.
\ee
This could be a strong inequality if the $\mu$ term in Eq.~(\ref{Friedmann2}) is dominated by the matter energy density.
In any case from Eq.~(\ref{tildeH}) we see that
\be
\label{last}
-\left[1+ \left({16\pi\mu\over 3k}\right)^{1/2} \right]{d\phi\over d\tau}<\tilde H.
\ee

It is clear from Eqs.~(\ref{a0dot}) and (\ref{last}) that for any choice of $\mu$--so long as it is positive,
\be
\Big|{d\mathfrak{a}_0/d\tau\over \mathfrak{a}_0}\Big| < \tilde H = {d\tilde b/d\tau\over \tilde b}.
\ee
Thus within any reasonable TeVeS theory $\mathfrak{a}_0$'s evolution is slower than the Hubble expansion at the same epoch, and it can be much slower, provided only $\mu$ is not small compared to unity in recent epochs and $k<1$.

A case in point is the TeVeS theory investigated in detail in Ref.~\onlinecite{BekPRD}.  It incorporates a function $\mathcal{F}(y)$ for which $\mu(y)>1$ for $y<0$.  As shown there one then needs $k\ll 1$ for TeVeS cosmology to be consistent with causality.  Thus by Eq.~(\ref{last}) $|d\phi/d\tau|\ll \tilde H$.  Then by virtue of Eq.~(\ref{a0dot}) this implies
\be
\label{a0dotnew}
\Big|{d\mathfrak{a}_0/d\tau\over \mathfrak{a}_0}\Big| \ll \tilde H = {d\tilde b/d\tau\over \tilde b}.
\ee
Thus at all epochs the evolution of $\mathfrak{a}_0$ occurs on a timescale much longer than Hubble's. Put another way, as one goes back in time, $\mathfrak{a}_0(z)$ grows much slower than $\tilde b_0/\tilde b(z)$ or $1+z$.

\section{conclusions}
\label{conc}

In this work we have calculated Newton's constant $G_N$ and the MOND acceleration scale $\mathfrak{a}_0$ in terms of TeVeS'  parameters.  We find that $G_N$ does not depend on the dynamical scalar field of the theory, and is thus strictly constant in cosmology.   This corrects an impression that one might obtain from Ref.~\onlinecite{BekPRD}.   It also shows that analogies drawn between TeVeS and familiar scalar-tensor theories  can lead to incorrect inferences.   Our result agrees with known facts: all existing data point to a nonvarying $G_N$~\cite{copi:171301}.     

We also find here that in a cosmological setting $\mathfrak{a}_0$  varies as the exponential of the scalar field, thus decreasing with time.  However, a detailed consideration of TeVeS isotropic cosmological models strongly suggests that the $\mathfrak{a}_0$ variation occurs on scales much longer than the Hubble scale.  This result is in contrast to a naive view which regards $\mathfrak{a}_0$ as physically connected to the Hubble parameter; in such eventuality $\mathfrak{a}_0$ variation would most likely occur on the Hubble scale (we have discussed inevitable ambiguities in this point of view).  

At present there are not enough quality data to test the TeVeS prediction of  slow $\mathfrak{a}_0$ evolution.   Clues as to the evolution of $\mathfrak{a}_0$ could be gleaned from existing data on the Tully-Fisher relation at epoch $z\sim 1$. The Tully-Fisher relation in the form $v_{\infty}^4=G_N\mathfrak{a}_0 M$, with $M$ the total baryonic mass of the galaxy and $v_{\infty}$ its
asymptotic rotation velocity, emerges naturally in MOND.  Evolution of $\mathfrak{a}_0$ would entail  evolution of the coefficient in the Tully-Fisher relation or, equivalently, of the zero point of the plot  of $\log M$ vs $\log v_\infty$ for  disk galaxies,  The meager available data are consistent with no evolution of the Tully-Fisher relation back  to $z\approx 0.6$~\cite{Flores:intermediate-z}.  In addition,  Milgrom's MOND analysis~\cite{milgreview2} of recent data by Genzel, Tacconi et al.~\cite{Genzel} on the rotation curve of
a galaxy at $z=2.38$ seems to be consistent with unchanging $\mathfrak{a}_0$ (although that rotation curve does not extend as far as would be desired for this kind of an inference). 

\begin{acknowledgments}
We thank Mordehai Milgrom for a critical reading and useful suggestions.  This research was supported by grant  694/04  of the Israel Science Foundation, established by the Israel Academy of Sciences and Humanities.
\end{acknowledgments}

\appendix

\section{Calculation of $G_N$ in Brans-Dicke theory}

We show here that the methodology of Sec.~\ref{sec:Newton} will yield familiar results when applied to a pure scalar-tensor theory such as Brans-Dicke (BD) theory.  Whenever feasible we shall couch the equations in the notation of Sec.~\ref{sec:Newton}.

Following Dicke~\cite{dicke} we transform the BD gravitational action~\cite{brans-dicke} to the Einstein frame; we shall, however, leave the matter action in the physical frame in parallel with our treatment of TeVeS~\cite{BekPRD}:
\begin{eqnarray}
S&=& {1\over 16\pi G}\int\left[R-{\scriptstyle 1\over \scriptstyle 2}(2\omega+3){\lambda_{,\alpha} \lambda,^\alpha\over\lambda^2}\right](-g)^{1/2} d^4 x
\nonumber
\\
&+&\int \mathcal{L}_{\rm m}(-\tilde g)^{1/2} d^4 x
\end{eqnarray}
 In the above $\omega$ is the celebrated BD parameter and  $\lambda$, a dimensionless entity,  represents the BD field in units of the fundamental constant $G^{-1}$, i.e., $\lambda=G\phi$.  The first line of the action is stated in terms of $g_{\alpha\beta}$ while the matter action takes its usual form when written in the $\tilde g_{\alpha\beta}$ metric.  In BD theory 
\be
\label{iden}
\tilde g_{\alpha\beta}=\lambda^{-1} g_{\alpha\beta};\quad \tilde g^{\alpha\beta}=\lambda g^{\alpha\beta};\quad (-\tilde g)^{1/2}=\lambda^{-2} (-g)^{1/2}.\ee

On account of the definition of the matter's energy momentum tensor as a variational derivative of the matter action we have
\begin{eqnarray}
-2\,\delta [\mathcal{L}_{\rm m}(-\tilde g)^{1/2}] &=&\tilde T_{\alpha\beta}(-\tilde g)^{1/2}\delta\tilde g^{\alpha\beta}
\nonumber
\\
=(- g)^{1/2}\big[\tilde T_{\alpha\beta}\,\delta g^{\alpha\beta}\lambda^{-1}&+&\tilde T_{\alpha\beta}\,\tilde g^{\alpha\beta}\lambda^{-3}\delta\lambda\big],
\end{eqnarray}
where the second line results on account of the transformations (\ref{iden}).
Now variation of $g_{\alpha\beta}$ in $S$ together with the identity
\be
\delta [ R (-g)^{1/2}]=G_{\alpha\beta}(- g)^{1/2}\delta g^{\alpha\beta}+\textrm{boundary terms}
\ee
and our last result yields the gravitational equations
\begin{eqnarray}
G_{\alpha\beta}&=&(2\omega+3)\lambda^{-2}\left[\lambda_{,\alpha}\lambda_{,\beta}-{\scriptstyle 1\over \scriptstyle 2}\lambda_{,\mu}\lambda_,^\mu g_{\alpha\beta} \right]\nonumber
\\
&+&8\pi G \tilde T_{\alpha\beta}\lambda^{-1},\label{BDeq}
\end{eqnarray}
the counterpart of our Eqs.~(\ref{metric_eq}),
while variation with respect to $\ln\lambda$ yields the BD scalar equation in the form
\be
\label{scalareqBD}
{1\over (-g)^{1/2}}[g^{\alpha\beta}(\ln\lambda)_{,\alpha}(-g)^{1/2}]_{,\beta}
={8\pi G\over 2\omega +3}\tilde T_{\alpha\beta}\,\tilde g^{\alpha\beta}\lambda^{-2},
\ee
which is the counterpart of Eq.~(\ref{scalar_eq}).

Let us solve the equations for a stationary situation to linear order by writing
\be
g_{\alpha\beta}=\eta_{\alpha\beta}+h_{\alpha\beta}\quad\textrm{and}  \quad \lambda=\lambda_c+\zeta,
\ee
with $\eta_{\alpha\beta}$ and $\lambda_c$ the asymptotic values of the Einstein metric and scalar field (where spacetime is assumed flat).
Now according to Eq.~(\ref{matterT}) for ideal fluid matter $\tilde T_{\alpha\beta}\,\tilde g^{\alpha\beta}=-\tilde\rho+3\tilde p$.  In first approximation we may neglect the $\tilde p$.  Then to first order in $\zeta$ and $h_{\alpha\beta}$ (and neglecting any temporal variation of the cosmological boundary value $\lambda_c$) Eq.~(\ref{scalareqBD}) takes the form
\be
\nabla^2\zeta = -{8\pi G/\lambda_c\over 2\omega+3}\tilde\rho,
\ee 
whence in analogy with Eq.~(\ref{TeVeSV}) 
\be
\zeta={2 G/\lambda_c\over 2\omega+3}\int\frac{\tilde{\rho}(\mathbf{x}')}{|\mathbf{x}'-\mathbf{x}|}d^3x'.
\ee

This last result shows that the first term in the  r.h.s. of Eqs.~(\ref{BDeq}) is of second order in $G\tilde\rho$, and thus negligible compared to the matter term. Again, from Eq.~(\ref{matterT}) we see that here $T_{\alpha\beta}\approx\tilde\rho v_\alpha v_\beta$.  If the matter is static we have from the normalization of $v_\alpha$ in the physical frame that $v_\alpha v_\beta=-\tilde g_{tt}\,\delta_\alpha^t\delta_\beta^t$.  This will also be true to good approximation if the matter flows in space provided that $v_\alpha$'s spatial part $\mathbf{v}$ is small compared to unity (errors will be of $\mathcal{O}(\mathbf{v}^2)$).  Hence the BD gravitational  equations are
\be
G_{\alpha\beta}\approx -8\pi G \lambda^{-1}\tilde\rho\,  \tilde g_{tt}\,\delta_\alpha^t\delta_\beta^t \approx 8\pi G \lambda_c^{-2}\tilde\rho  \,\delta_\alpha^t\delta_\beta^t,
\ee
where we have used relations (\ref{iden}) and dropped from the last expression. subdominant terms with extra factors of $h_{\alpha\beta}$ and $\zeta$.
These equations are just the GR Einstein equations in metric $g_{\alpha\beta}$ for a quasistatic mass-energy distribution $\tilde\rho$, but  with $G/\lambda_c^{-2}$ playing the role of gravitational constant.  

We know that to first order such Einstein equations have the line element (\ref{linearEE}) as solution with $V$ signifying the usual Newtonian potential.  In light of our remark about the gravity constant  we must write here the following analog of Eq.~(\ref{TeVeSV}):
\be
V=-G\lambda_c^{-2}\int\frac{\tilde{\rho}(\mathbf{x}')}{|\mathbf{x}'-\mathbf{x}|}d^3x'.
\ee
Using the transformations (\ref{iden}) we evidently have to first order in $V$ and $\zeta$ that
\begin{eqnarray}
d\tilde s^2&=&-(1+2V-\zeta/\lambda_c)\lambda_c^{-1} dt^2
\nonumber
\\
&+&(1-2V-\zeta/\lambda_c)\lambda_c^{-1}d{\mathbf{x}}\cdot
d{\mathbf{x}}
\end{eqnarray}
In order that the physical line element be asymptotically Minkowski, we must define, in analogy with relations~(\ref{coordrel}), the physical time $\tau$ and physical lenght coordinates $\tilde{\mathbf{x}}$: 
\be
\tilde{\mathbf{x}}=\lambda_c^{-1/2}\mathbf{x}
;\qquad \tau=\lambda_c^{-1/2} t.
\ee
The line element here has thus a form that contrasts that of Eq.~(\ref{linear}) for GR and TeVeS:
\be
d\tilde{s}^2=-(1+2\Phi_N)d\tau^2+(1-2\kappa \Phi_N)d\tilde{\mathbf{x}}\cdot
d\tilde{\mathbf{x}}.
\ee
Here
\begin{eqnarray}
\Phi_N&=&V-{\scriptstyle 1\over \scriptstyle 2}\zeta/\lambda_c=-G_N\int\frac{\tilde{\rho}(\tilde{\mathbf{x}}')}{|\tilde{\mathbf{x}}'-\tilde{\mathbf{x}}|}d^3\tilde x',
\label{BDPhiN}
\\
G_N&=&{G\over\lambda_c}{2\omega+4\over 2\omega+3},
\\
\kappa&=&{\omega+1\over \omega+2}.
\end{eqnarray}
In Eq.~(\ref{BDPhiN}) we have absorbed one factor $\lambda_c^{-1}$ into the integral to convert from $\mathbf{x}$ to $\tilde{\mathbf{x}}$.

Comparison with Eqs.~(\ref{NewtPot})-(\ref{linear}) shows that $G_N$ here is properly regarded as the Newtonian gravity constant.  Our $G_N$ concurs with Brans and Dicke's~\cite{brans-dicke} showing clearly that in BD theory the Newtonian gravity ``constant'', by virtue of its strong $\lambda$ dependence, evolves cosmologically, in contrast to the case of GR or of TeVeS.  Our value for the coefficient $\kappa$ also coincides with that obtained by Brans and Dicke~\cite{brans-dicke}.  The fact that $\kappa\neq 1$ is responsible for gravitational lensing being smaller in BD theory than in GR, is again in contrast to the TeVeS case for which gravitational lensing is the same as that in GR for the same source $\tilde\rho(\tilde{\mathbf{x}})$.

\end{document}